\documentclass{aa}
\usepackage{graphics}
\input{psfig}
\begin{document}

\thesaurus{08(08.02.4, 08.02.6, 08.06.3, 08.09.2)}

\title{A calibration of $\iota$\,Pegasi system}

\author{P. Morel, Ch. Morel, J. Provost and G. Berthomieu}

\institute{
D\'epartement Cassini, UMR CNRS 6529, Observatoire de la C\^ote 
d'Azur, BP 4229, 06304 Nice CEDEX 4, France
}

\offprints{P. Morel}
\mail{Pierre.Morel@obs-nice.fr}

\date{Received date / Accepted date}

\maketitle

\begin{abstract}
Recent observations provide determinations of
individual masses, chemical composition and metallicity
of the components of the spectroscopic and interferometric
binary $\iota$\,Peg (Boden et al. \cite{bkb98}). Using updated physics, to
calibrate the system, we have computed using the stellar
evolutionary code CESAM (Morel \cite{m97}),
evolutionary sequences of stellar models with the masses of $\iota$\,Peg\,A
 $1.326\,M_\odot$ and $\iota$\,Peg\,B $0.819\,M_\odot$ (Boden et al.
 $loc.\,cit$) and with different values of the
mixing-length parameter $\alpha$, the helium $Y$ and the
heavy element $Z$ initial
mass fraction with the constraint of the observed metallicity. Adopting
effective temperatures and luminosities, as derived from observations with the
bolometric corrections, and the empirical scale of
temperatures of Alonso et al. (\cite{aam95}, \cite{aam96}),
we find $\alpha_{\rm A}=1.46$, $\alpha_{\rm B}=1.36$, $Y=0.278$, $Z=0.017$.
The evolution time, including pre-main
sequence, is found within $\sim40\,{\rm My}\la t_{\rm ev}\la 0.5\,{\rm Gy}$.
The calibrated models of $\iota$\,Peg.\,A and B are non homogeneous zero age main
sequence stars with the evolutionary time $t_{\rm ev}=56$\,My.
Due to the large uncertainties of their determinations,
the values derived for the mixing-length parameters are smaller than the solar
one but however marginally compatible with it.
Our results ought to be improved as soon as a more accurate value of
the magnitude difference in the V filter will be available.
Detailed spectroscopic analysis for both components looks practicable,
so it is urgently needed.

\end{abstract}

\section{Introduction}\label{sec:int}
Along the last decade, owing to significant
improvements of our knowledge of fundamental stellar data, trigonometric
parallaxes with HIPPARCOS (ESA \cite{e97}), 
spectroscopic-speckle mass determination (e.g. Scarfe et al. \cite{s94}),
stellar diameters
(e.g. Koechlin \& Rabbia \cite{kr85}, Di\,Benetto \& Bonneau \cite{db91}),
photometric 
(Mermillod et al. \cite{mmh97}) and spectroscopic data (Cayrel de Strobel et
al. \cite{c97}), the constraints on stellar modeling
became more stringent leading to improvements of stellar evolution theory.
Meanwhile, physics, e.g.
 opacities (Iglesias \& Rogers \cite{ir96}), equation of state
(Rogers et al. \cite{rsi96}; D\"appen \cite{d96}),
thermonuclear reaction rates (Adelberger et al. \cite{a98}, Angulo et al.
\cite{a99}), have been improved. Most of these new data are nowadays
available in free access in data banks.

With the exception of the Sun, modeling a single star is not a closed
problem because the number of indeterminate parameters is larger than the
observed ones.
For some well known binaries, e.g. $\alpha$\,Cen
(Demarque et al. \cite{dga86}; Noels et al. \cite{n91}), 
with the reasonable hypothesis of a common origin for both components, i.e. same
initial chemical composition and age, it is possible to confront observations
and theoretical stellar evolution sequences. One can derive
estimates for age, helium content and metallicity,
fundamental quantities for our understanding of the galactic chemical
evolution.
One derives also the mixing-length parameter $\alpha$ -- also named ``convection
parameter''.

For a {\em fixed} physics, the value of the 
mixing-length parameter $1.0\loa \alpha_\odot\loa 2.5$ 
derived from a calibrated solar model 
is currently used to model other stars. But, one can
ask why the value adjusted for the present Sun should be applied to stars
with different masses, ages and/or chemical composition.
Fixing the physics, Fernandes et al. (\cite{flbm98}) have
shown that $\alpha_\odot$
is valuable for a sample of low mass binaries
$0.6\,M_\odot\la M_\star\la 1.0\,M_\odot$
in the metallicity range $\rm -0.65\la [Fe/H]\la +0.0$. Indeed,
for $\alpha$\,Cen Fernandes \&
Neuforge (\cite{fn95}) obtained a value close to the solar one.
More recently, Pourbaix et al.
(\cite{pnn99}) using improved mass and gravity determinations for both
components of $\alpha$\,Cen were able to 
relax the assumption of a unique value for the convection parameter. They
obtained $\alpha_{\rm A}=1.86\pm0.80$ ($resp.$ $\alpha_{\rm B}=2.1\pm0.80$)
for the A ($resp.$ B) component. Due to the range of uncertainties,
this result does
not rule out the assumption of a constant value for the mixing-length
parameter for masses and metallicities smaller
or slightly larger to solar values.

Up to now, the calibrations of double stars have especially
concerned main sequence 
stars with ages of few Gy.
As soon as new data become available, it is important to extend 
the sample of calibrated binaries. In most cases the individual masses are
inaccurately known and the isochrones are not really constrained. The
spectroscopic and interferometric binary
$\iota$\,Peg. is a particularly interesting object since the two masses are
accurately known (Boden et al. $loc.\,cit.$). Moreover,
the mass and the metallicity of the primary are
slightly larger than solar. It is a ``solar like'' star whose
seismological properties have been studied (e.g., Houdek et al.
\cite{h95}, Houdek \cite{h96}, Michel et al. \cite{mh99}).
Some stars of this group are COROT targets (Baglin \cite{bc98}).
The values of the mixing-length parameters, as derived from calibration
 for $\iota$\,Peg.\,A and B,
will provide a test for the 2-D numerical  calculations of compressible
convection (Ludwig et al. \cite{lfs99}).

Some authors have already derived estimates for the age of the $\iota$\,Peg.
binary system. A value of $\sim 80$\,My was derived
from the observed lithium depletion
(Fekel \& Tomkin \cite{ft83}, Gray \cite{g84}).
Lyubimkov et al. \cite{lpr91} have obtained the slightly larger value of
 $170\pm80$\,My, from comparison with evolutionary models. None of these
 estimates is based on evolutionary models taking explicitly into account,
as we do, the microscopic
 diffusion and gravitational settling, basic phenomena in stellar modeling.

In this paper we report an attempt to calibrate $\iota$\,Peg. using stellar
evolution models including pre-main sequence and microscopic diffusion.
It is organized as follows:
In Sect.~\ref{sec:ipeg} we collect and discuss the observations.
Section~\ref{sec:tl} is devoted to derivations of
effective temperatures and luminosities.
The method of calibration is described in Sect.~\ref{sec:cal}.
Section~\ref{sec:comp} presents the stellar modeling procedure. In
Sect.~\ref{sec:res} we give the results with emphasis on a seismological
analysis. We discuss the results and conclude in Sect.~\ref{sec:dis}.

\begin{table}
\caption[]{
Astrometric, photometric and spectroscopic adopted data of $\iota$\,Peg.
 $m_{\rm V}$, $\Delta m_{\rm V}$
($resp.$ $m_{\rm K}$, $\Delta m_{\rm K}$) are the global apparent
magnitudes and magnitude
differences in the V ($resp.$ K) band. 
$\varpi_{\rm orb}$ is the orbital parallax. $\cal M_{\rm A}$,
 $\cal M_{\rm B}$ are the masses,
$\rm [Li]_A$, $\rm [Li]_B$ the lithium abundances (H=12),
of A and B components, respectively. $\rm [Fe/H]$ is the
metallicity. The flag ``$^\ddag$'' signals a value estimated by
{\em ourselves}.
}\label{tab:data}
\begin{tabular}{lllllll} \\  \hline \\
Spectra	           &F5V -- G8V		& Fekel \& Tomkin (\cite{ft83})\\
$m_{\rm V}$	   &$3.771\pm(0.001)^\ddag$     & Mermillod et al. (\cite{mmh97})\\
$\Delta m_{\rm V}$ &$2.68\pm(0.05)^\ddag$     & Fekel \& Tomkin (\cite{ft83})\\            
$m_{\rm K}$        &$2.656\pm0.002$     & Bouchet et al. (\cite{b91})\\   
$\Delta m_{\rm K}$ &$1.610\pm0.021$     & Boden et al. (\cite{bkb98})\\
$\varpi_{\rm orb}$    &$86.91\pm1.0$\,mas    & Boden et al. (\cite{bkb98})\\
$\cal M_{\rm A}$   &$1.326\pm0.016\,M_\odot$ & Boden et al. (\cite{bkb98})\\
$\cal M_{\rm B}$   &$0.819\pm0.009\,M_\odot$ & Boden et al. (\cite{bkb98})\\
$\rm [Li]_A$    &$2.9\pm(0.4)^\ddag$\,dex& Fekel \& Tomkin (\cite{ft83})\\
$\rm [Li]_B$    &$2.6\pm 0.6$\,dex& Fekel \& Tomkin (\cite{ft83})\\
$\rm [Fe/H]$       &$(+0.00\pm0.1)^\ddag$\,dex	& See text\\
\\ \hline \\
\end{tabular}
\end{table}
\section{Observations of $\iota$\,Peg.}\label{sec:ipeg}
$\iota$\,Pegasi (HR 8430, HD 210027) is a nearby ($d\simeq11.75$\,pc), a
wonderful double lined spectroscopic and interferometric
double star. The binarity was
discovered more than a century ago by Campbell (\cite{c899}), who notices
``The velocity of
$\iota$\,Pegasi in the line of sight is variable''.
Table~\ref{tab:data} shows the astrometric, photometric and
spectroscopic observational data retained for the present investigation.

\subsection{Orbit, masses, parallax and diameters}
Recently, Boden et al. ($loc.\,cit.$) using
observations of the Palomar Testbed Interferometer
computed the visual orbit of $\iota$\,Peg. and derived the individual masses.
The orbital period is accurately known $p=10.213033\pm1.3\,10^{-5}$\,day
(Fekel \& Tomkin \cite{ft83}).
The spectroscopic and visual orbits for $\iota$\,Peg. are each statically 
consistent with being circular. That corroborates the result of
Zahn \& Bouchet (\cite{zb89}) who predict, for close binary systems
with masses ranging from $0.5$ to $1.25\,M_\odot$, that the
orbit is circularized during the Hayashi phase.
Moreover, since the period of $\iota$\,Peg. is 
larger than $\approx 8$\,days, it is expected
that the main sequence is reached with
components rotating faster than the orbital rate,
Zahn \& Bouchet ($loc.\,cit.$).

The orbital parallax derived by Boden et al.
($loc.\,cit.$), as seen in Table~\ref{tab:data}, is in close agreement
with the trigonometric $\varpi_{\rm trig}=85.06\pm0.71$\,mas (ESA \cite{e97})
determination by HIPPARCOS. Though
 the accuracy of the HIPPARCOS parallax is slightly higher, we have
used the Boden et al. value for global consistency.
Boden et al. ($loc.\ cit.$), 
assuming two different limb darkening laws, have obtained consistent
estimates of diameters for each component:
$\diameter_{\rm A}=0.99\pm0.05$ mas,
$\diameter_{\rm B}=0.70\pm0.10$ mas.
The illuminating Fig. 1 of Boden et al. ($loc.\ cit.$) shows that
the previous values
of diameters are not large enough for eclipsing,
despite the high inclination $i=95.85^\circ\pm 0.22^\circ$ and
the small semi-major axis $a=10.33\pm0.10$\,\,mas.
From the previous interferometric data and
orbital parallax one derives the ``interferometric'' radii of
$R_{\rm A\,in}=1.224\pm0.025\,R_\odot$ and
$R_{\rm B\,in}=0.872\pm0.050\,R_\odot$,
respectively, for primary and secondary.
As emphasized by Boden et al. ($loc.\ cit.$), the sum of the radii is
consistent with the  absence of evidence of photometric eclipse.
For the gravities one obtains:
$\log_{10}\,g_{\rm A\,in}=4.39\pm0.23$ and
$\log_{10}\,g_{\rm B\,in}=4.47\pm0.65$.

\subsection{Spectroscopy}\label{sec:sp}
From analyses of 2.5\,\AA\ per mm CCD spectra,
Lyubimkov et al. (\cite{lpr91}) have derived effective temperatures
$T_{\rm eff\,A\,sp}=6750\pm150$\,K,
$T_{\rm eff\,B\,sp}=5350\pm350$\,K, and gravities
$\log_{10}\,g_{\rm A\,sp}=4.35\pm0.05$,
$\log_{10}\,g_{\rm B\,sp}=4.57\pm0.10$. With the values
of the masses derived by Boden et al. ($loc.\,cit.$),
that correspond to a ``spectroscopic'' luminosity of
$L_{\rm A\,sp}=3.024\pm0.62\,L_\odot$ and
$L_{\rm B\,sp}=0.444\pm0.219\,L_\odot$,
respectively for the primary and secondary,
and to radii
$R_{\rm A\,sp}=1.274\pm0.07\,R_\odot$ and 
$R_{\rm B\,sp}=0.772\pm0.089\,R_\odot$.
Another effective
temperature value $T_{\rm eff}=6488$\,K, derived from spectroscopic analysis,
was just recently listed by Boesgaard et al. (\cite{b99}).
Note that the constraint on the diameters, due to the absence of evidence of
photometric eclipse (Boden et al. $loc\,cit.$), is also fulfilled with the
``spectroscopic'' values of the radii.

\subsection{Chemical composition}\label{sec:cp}
Lyubimkov et al. (\cite{lpr91}) have qualified the chemical composition
as ``close to normal''. The Catalog of [Fe/H] (Cayrel de Strobel
\cite{c97}) recommends $\rm [\frac {Fe}H]=+0.10$. Another value of the metallicity
$\rm [\frac {Fe}H]=-0.13$ is due to Duncan (\cite{d81}). According to 
Boesgaard et al. (\cite{b99}) the metallicity is $\rm [\frac {Fe}H]=-0.08$.
Faced with these rather scattered data, we have adopted for our calculations
 $\rm [\frac {Fe}H]=0.00\pm0.10$ (Cayrel de Strobel, private communication).
Except for lithium (see next), none of the authors has reported
significant differences between the surface chemical composition of primary
and secondary. It is an indication that both components of $\iota$\,Peg. are
newly formed stars, since the gravitational settling, which
acts more efficiently in a $1.326\,M_\odot$
than in a $0.819\,M_\odot$, still has not have enough time to differentiate
significantly the surface chemical composition of primary and secondary
(e.g. Morel \& Baglin \cite{mb99}).
Therefore, the initial metallicity of $\iota$\,Peg.
is certainly very close to the nowadays
observed value.

Table~\ref{tab:data} shows the adopted
values of lithium abundances. They are in agreement
with the measurements of
Lyubimkov et al. (\cite{lpr91}): $\rm [Li]_A=3.25$ for primary and
$\rm [Li]_B=2.58$ for secondary. 
According to the standard evolution theory, for an isolated star
of solar metallicity and
less massive than $M_\star\loa 0.85\,M_\odot$, all lithium is
destroyed at the end of the
fully convective phase of the pre-main sequence.
That results from how fast the convective zone recedes from the core where the
lithium is depleted -- recall that lithium burns
as soon as the temperature
is greater than $T\goa 2.4$\,MK. In the core of a late type star, the
increase of temperature is
slow, so are the decreases of opacity and radiative gradient. Then,
the fully convective state lasts longer than in a more massive star.
With respect to an isolated star, $\iota$\,Peg.\,B is lithium under-depleted.
For not isolated stars
under-depletion can occur on the main-sequence in binary systems
which are tidally locked for orbital periods below $\approx 8$\,days
(Zahn \cite{z94}). The period of $\iota$\,Peg. (10 days)
is slightly larger than this limit, 
as supported by the no-synchronous rotation
found by Fekel \& Tomkin (\cite{ft83}). But,
Gray (\cite{g84}) deduced, from the analysis of higher resolution spectra,
that primary and secondary are slow rotators
$v_{\rm A}\sin i=6.5\pm0.3$\,km\,s$^{-1}$,
$v_{\rm B}\sin i=5.\pm1$\,km\,s$^{-1}$, and
argued that both components are in synchronous rotation.
As previously conjectured, the components may have
reached the main sequence with
rotation velocities larger than the orbital one and the individual
rotation velocities may have decreased, the system being again tidally locked.
Figure~\ref{fig:AB} shows that the locus of $\iota$\,Peg.\,B 
in the HR diagram excludes the
possibility that the secondary is nowadays still in the pre-main
sequence, before the lithium burning phase which occurs soon after
the end of Hayashi's track.
Therefore, the under-depletion of lithium, observed in $\iota$\,Peg.\,B, is
unexplained. There is an alternative  
either, i) along the pre-main sequence, for some unknown reason,
the mixing of the convection zone is
slowed and lithium is not destroyed or, ii) at
the beginning of the main sequence, coming from somewhere,
lithium has been provided to the outer
convection zone of the secondary. Due to the distance between the components
$a\approx0.12$\,AU there is no possibility for mass exchange between the
components (F. van\,'t Veer, private communication). Then the first part
(i) of the alternative appears to be the most reliable.
That reinforces the conjecture
that $\iota$\,Peg. is certainly composed of two newly formed stars.
In our calculations, to mimic the unknown process leading to the
under-depletion of lithium in the secondary, the
convective zones will be only mildly mixed
in the course of the pre-main sequence (see Sect.~\ref{sec:mix}).

\subsection{Photometry}\label{sec:ph}
 $\iota$\,Peg.\,A is a MK standard and its photometry is accurately known.
 Another value of the K apparent magnitude $m_{\rm K}=2.623\pm0.016$ is due to
 Carrasco et al. (\cite{c91}). For the magnitude difference $\Delta m_{\rm V}$
 in the V band we have adopted the estimate of 
Fekel \& Tomkin (\cite{ft83}). Their analysis is based on
spectroscopic and photometric arguments.
Comparing the strength of iron lines of $\iota$\,Peg.\,B and 
$\epsilon$\,Eri, a K2V star, they derive, for $\iota$\,Peg.
 an estimate of the magnitude difference
$\Delta m_{\rm R}=2.5\pm0.3$ in the R band.
Such a value of $\Delta m_{\rm R}$ characterizes
stars between G5V and K0V. Then, they deduce $\Delta m_{\rm V}=2.68$ from
the sensitivity of the global K magnitude for binary systems composed of a 
F5V primary and secondaries with spectral types between G5V and K0V.
Fekel \& Tomkin ($loc.\ cit.$) do not give the uncertainty
$\delta\Delta m_{\rm V}$ of
their estimation of $\Delta m_{\rm V}$;
it is difficult to derive such value from their subtle analysis.
We have {\em estimated} $\delta\Delta m_{\rm V}=\pm0.05$.
This quantity is the most poorly known parameter of our calibration.
Lyubimkov et al. (\cite{lpr91}) have
reported the estimate $\Delta m_{\rm V}=2.13$ which significantly differs from
the value derived by Fekel \& Tomkin ($loc.\ cit.$).

\begin{table}
\caption[]{
Physical parameters derived from data of Table~\ref{tab:data} using
$\varpi_{\rm orb}$.
$Bc_{\rm K}$ is the bolometric correction relative to the K magnitude.
$M_{\rm bol}$, $M_{\rm V}$ ($resp.$ $M_{\rm K}$) are respectively
the bolometric and absolute magnitudes in the V ($resp.$ K) band.
}\label{tab:lt}
\begin{tabular}{llllll} \\  \hline \\
Parameter          &$\iota$\,Peg.\,A &$\iota$\,Peg.\,B\\ \\ \hline \\
$Bc_{\rm K}$       &$+0.97$	    &$+1.82$ \\
$M_{\rm bol}$      &$3.54\pm0.08$   &$6.01\pm0.09$     \\
$M_{\rm V}$        &$3.56\pm0.03$   &$6.24\pm0.07$     \\
$M_{\rm K}$        &$2.57\pm0.03$   &$4.18\pm0.03$ \\
$T_{\rm eff}$      &$6642\pm52$\,K  &$4992\pm115$\,K   \\
$L/L_\odot$        &$3.041\pm0.23$ &$0.3145\pm0.027$  \\
 \\ \hline \\
\end{tabular}
\end{table}

\section{Luminosities and effective temperatures}\label{sec:tl}
Faced with the scattered values of luminosities and effective temperatures,
derived either from spectroscopic
(Sect.~\ref{sec:sp}) or photometric (Sect.~\ref{sec:ph}) observations,
we make the choice to use the photometric data alone.
Table~\ref{tab:lt} gives the physical parameters of primary and secondary as
derived from data of Table~\ref{tab:data}.
The bolometric corrections for the K band
are obtained using the bolometric corrections of Alonso et
al. (\cite{aam95}) with a correction of the zero point giving a bolometric
magnitude of 4.75 to the Sun (Cayrel, private communication).
The uncertainty of the method
is of the order of $\approx\pm0.05$ magnitude.
The effective temperatures are derived from the Alonso et
al. (\cite{aam96}) empirical scale of temperature, valid
for the low main sequence
from F0V to K0V. The uncertainty of the fit is of
the order of $\widetilde{\delta T_{\rm eff}}\approx\pm30$\,K.
We are aware that
many tidally locked binary systems, having a lithium under-depleted
late type component, are recognized as chromospherically active
(Barrado y Navascu\'es et al. \cite{b97}). As 
$\iota$\,Peg. {\em is not} identified as
chromospherically active (Barrado y Navascu\'es et al. $loc.\ cit.$)
the use of the Alonso et al. (\cite{aam95}, \cite{aam96})
procedures is valid. From the previous photometric and
orbital data one derives the ``photometric'' radii of
$R_{\rm A\,ph}=1.319\pm0.023R_\odot$ for primary and
$R_{\rm B\,ph}=0.751\pm0.027R_\odot$ for secondary.
Indeed, for the gravities one obtains:
$\log_{10}\,g_{\rm A\,ph}=4.32\pm0.02$,
$\log_{10}\,g_{\rm B\,ph}=4.60\pm0.03$, values in agreement with
their ``spectroscopic'' determination of Sect.~\ref{sec:sp}.
Note that the constraint on the diameters, due to the absence of evidence of
photometric eclipse (Boden et al. $loc\,cit.$), is also fulfilled with the
``photometric'' values of radii.

The effective temperatures, luminosities and radii of
 F5V and G8V standard stars are respectively 6440\,K, 3.20\,$L_\odot$,
 $1.44\,R_\odot$
and 5570\,K, 0.66\,$L_\odot$, $0.874\,R_\odot$ (Schmidt-Kaler \cite{sk82}).
According to radii and luminosities, derived previously either
from photometric data or from
spectroscopic data (Sect.~\ref{sec:sp}),
both components of $\iota$\,Peg. are under luminous
and have smaller diameters than standards. That reinforces the
conjecture of a small age. $\iota$\,Peg. being certainly a young 
binary system, the models have to take properly into account the
pre-main sequence evolution.

\paragraph{Uncertainty domains.}
For each component the luminosity and the effective temperature, as derived
from Alonso et al. ($loc.\ cit.$) semi empirical adjustments, are functions of
absolute magnitudes $M_{\rm V}$ and $M_{\rm K}$. Then, the
luminosity and the effective temperature have the following
{\em non-linear} functional
dependency with respect to the observed parameters:
\begin{eqnarray}
L&=&L(m_{\rm V},\Delta m_{\rm V},m_{\rm K},
\Delta m_{\rm K},\varpi,{\rm [Fe/H]}), \nonumber \\
T_{\rm eff}&=&T_{\rm eff}(m_{\rm V},\Delta m_{\rm V},m_{\rm K},
\Delta m_{\rm K},\varpi,{\rm [Fe/H]}). \nonumber
\end{eqnarray}
Only the uncertainties on the observed magnitudes, the distance and
the color index can be assumed
to be Gaussian. The statistical behavior of the metallicity determination,
of the estimated V magnitude difference (see Sect.~{\ref{sec:ph}) and of
the fits of Alonso et al. ($loc.\,cit.$) are unknown. Therefore, due to the
non-linear dependence of the effective temperature with respect to variables
some of which are correlated and not Gaussian, we have
estimated the uncertainty on the effective temperature using the sum:
\begin{eqnarray}
\Delta T_{\rm eff}&\approx&\left|\frac{\partial T_{\rm eff}}{\partial m_{\rm V}}\right|
\delta m_{\rm V}+
\left|\frac{\partial T_{\rm eff}}{\partial \Delta m_{\rm V}}\right|
\delta \Delta m_{\rm V}+\nonumber \\
&+&
\left|\frac{\partial T_{\rm eff}}{\partial m_{\rm K}}\right|
\delta m_{\rm K}+
\left|\frac{\partial T_{\rm eff}}{\partial \Delta m_{\rm K}}\right|
\delta \Delta m_{\rm K}+ \nonumber \\
&+&
\left|\frac{\partial T_{\rm eff}}{\partial \varpi}\right|
\delta \varpi+
\left|\frac{\partial T_{\rm eff}}{\partial {\rm [Fe/H]}}\right|
\delta {\rm [Fe/H]}+ \widetilde{\delta T_{\rm eff}}, \nonumber 
\end{eqnarray}
instead of the standard square root of the sum of the square of the standard
deviations.
A similar relation holds for the luminosity. Here, $\delta x$
is the uncertainty on $x$. Table~\ref{tab:lt} shows the
uncertainties derived in such a way for effective
temperatures and luminosities. They are maybe slightly optimistic but give a
reasonable order of magnitude.

\section{The method of calibration}\label{sec:cal}
Fixing the initial mass and the physics, the calculation
of a stellar model needs 4 unknowns namely,
the mixing-length parameter $\alpha$, the  helium $Y_0$ and
heavy elements $Z_0$ initial mass
fractions and the age. To model a binary,
assuming a common origin to each component,
the age and the initial chemical composition are identical for primary and
secondary. There are only $4\times 2-3=5$ unknowns.
There are 6 observables: luminosity,
effective temperature and metallicity for each component.
In case of $\iota$\,Peg., many presumptions of a small age
allow to assume that the observed
surface metallicity $(\frac ZX)_{\rm\iota\,Peg.}$
is the same for both components and then
close to its initial value (see Sect.~\ref{sec:cp}).
Therefore, there remain 5 observables and
5 unknowns: the mixing-length parameters
$\alpha_{\rm \iota\,Peg.\,A}$ and $\alpha_{\rm \iota\,Peg.\,B}$ of primary and
secondary, the initial helium $Y_0$ and metal $Z_0$ contents
and the age $t_{\rm \iota\,Peg.}$. 
As $1\equiv X+Y+Z$, the initial values of helium and metals are related by:
\[Z_0=\frac{(1-Y_0)}{1+(\frac ZX)_{\rm\iota\,Peg.}}
\times(\frac ZX)_{\rm\iota\,Peg.}.\]
This relationship holds {\em only} because the initial and present day
metallicities are alike for both components; otherwise $Y_0$ and $Z_0$
need to be derived independently.
Therefore, 4 calibration parameters remain:
\[\alpha_{\rm \iota\,Peg. A},\  \alpha_{\rm \iota\,Peg. B},
\ Y_0,\ t_{\rm \iota\,Peg.}.\] 
Since we have adopted
$[{\rm\frac{Fe}H}]_{\rm\iota\,Peg.}=0.0$, we used 
$(\frac ZX)_{\rm\iota\,Peg.}=(\frac ZX)_\odot=0.0245$
 (Grevesse \& Noels. \cite{gn93}).

\paragraph{Search of a solution.} In a first time we have explored
systematically the domain
of the calibration parameters. In a second step, more or
less empirically, we have refined the 
convection parameters and the initial helium mass fraction.
Doing that we are guided by two
facts, i) in the HR diagram an increase of the convection parameter moves
the evolutionary track towards larger temperatures at about
constant luminosity and, ii)
an increase of the helium content shifts the evolutionary track towards
larger effective temperatures {\em and} luminosities.
For stars with convective envelope, the first point (i)
results from the fact that
an increase of the efficiency of the convection deepens the base of the
convection zone and does not affect the core. The luminosity is then
 almost unchanged, while the adiabat of the external convection zone
 is adjusted on a larger
temperature value at the base of the convection zone, with the result of an
 increase of the effective temperature. The second point (ii) can be understood
 in the following way:
 an increase of the helium mass fraction leads to an increase of the mean
 molecular weight which, by turn, diminishes the pressure gradient.
 The quasi-static equilibrium is conserved via a temperature increase.
 A concomitant luminosity increase results from the enlargement of
 the nuclear energy generation magnified by the large power law
 dependence of nuclear reaction rates, with respect to temperature.
 In summary, in the HR diagram a change of $\alpha$ translates
 the locus of a model in effective temperature at constant luminosity,
 while one can adjust changes of $\alpha$
 {\em and} $Y$ in order to translate the locus of a model
 in luminosity at constant effective temperature.
   
Finally, the steepest descent algorithm
(e.g. Conte \& de\,Boor \cite{cb87}, Noels et al. \cite{n91}) is used
to improve the calibration parameters. As this gradient method converges
in the vicinity of {\em stable} solutions we have only consider
as a ``solution'' any
set of calibration parameters for which the gradient method converges.
We have kept for the {\em calibrated models} the solution which best fits
the central values of observational boxes in the HR diagram for each
component. Then we have estimated the domains of variations allowed to the
parameters in order that the locii of the models remain in these observational
boxes.

\section{The physics of the models}\label{sec:comp}
Basically the physics used in the models is the same as in
Morel et al. (\cite{mpb97}). The standard assumptions of stellar modeling are
made, i.e. spherical symmetry, no rotation, no magnetic field, no mass loss.
Each evolution is initialized with a homogeneous zero-age
pre-main sequence model in quasi-static contraction
with a temperature at center
close to the onset of the deuteron burning, i.e. $T_{\rm c}\sim0.5$\,MK.
We shall call the ``time of evolution'' of a model
the time $t_{\rm ev}$ elapsed from initialization. In a sequence of models
we shall designate by ``model of zero age main-sequence'' (ZAMS) the first
model, if any, where the nuclear
reactions dominate gravitation by more than 99\% as the primary energy source.
Typically ZAMS occurs after $\sim40$\,My of evolution for
$\iota$\,Peg.\,A and $\sim70$\,My for $\iota$\,Peg.\,B.

\subsection{Nuclear and diffusion network.}
The general nuclear network we used contains the following species:
\element[][1]{H},
\element[][2]{H},
\element[][3]{He},
\element[][4]{He},
\element[][7]{Li},
\element[][7]{Be},
\element[][12]{C},
\element[][13]{C},
\element[][14]{N},
\element[][15]{N},
\element[][16]{O},
\element[][17]{O} and 
\element[][]{Ex}; \element[][]{Ex} is an ``Extra'' fictitious
mean non-CNO heavy element with atomic mass 28 and charge 13
($\element[][]{Ex}\sim\element[][28]{Al}$) which
complements the mixture i.e.,
$X_{\element[][]{Ex}}=1-\sum_{i=\element[][1]{H}}^{\element[][17]{O}}X_i$
with $X_i$ as the mass fraction of the species labeled with
$i=\element[][1]{H},\ldots,\element[][17]{O}$.
With respect to time, due to microscopic diffusion processes, the
abundances of heavy elements are enhanced towards the center;  
\element[][]{Ex}  mimics that enhancement for the non CNO metals 
which contribute to changes of $Z$, then to opacity variations but
not to the nuclear burning and nuclear energy generation.
We have taken into account the 
most important nuclear reactions of PP+CNO cycles (Clayton, \cite {c68}).
The relevant nuclear reaction rates
are taken from the NACRE compilation (Angulo et al. \cite{a99}) with the
reaction
$\element[][7]{Be}(e^-,\nu\gamma)\element[][7]{Li}$ taken from the
compilation of
Adelberger et al. (\cite{a98}). Weak screening (Salpeter \cite{s54}) is assumed.
Owing to the fact that the nowadays observed surface chemical
composition is solar and close to the initial mixture (see Sect.~\ref{sec:cal}),
the initial fractions between the heavy elements within $Z$ are set to
their photospheric solar present day values. More details about this procedure
can be found in Morel \& Baglin (\cite{mb99}).
We have used the meteoritic value (Grevesse \& Sauval \cite{gs98}) for the
initial lithium abundance,
$\left[\element[][]{Li}\right]_0=3.31\pm0.04$ ($\rm
H\equiv12$).
For the calculations of the depletion, the lithium
is assumed to be in its most abundant isotope \element[][7]{Li} form.
The initial abundance of each isotope is derived from
isotopic fractions and initial values of
$Y\equiv \element[][3]{He}+\element[][4]{He}$ and $Z$ in order to fulfill 
the basic relationship $X+Y+Z\equiv1$ with
$X\equiv \element[][1]{H}+\element[][2]{H}$.
Microscopic diffusion is described by the simplified formalism of 
Michaud \& Proffitt (\cite{mp93}) with each  heavy element as a
trace element.

\subsection{Equation of state, opacities, convection and
atmosphere.}\label{sec:mix}
We have used the OPAL equation of state (Rogers et al. \cite{rsi96})
and the opacities of Iglesias \& Rogers (\cite{ir96})
for the solar mixture of Grevesse \& Noels (\cite{gn93}) 
complemented, at low temperatures by
Alexander \& Ferguson (\cite{af94}) opacities.

In the convection zones the temperature gradient is
computed according to the standard mixing-length theory
(B\"ohm-Vitense \cite{b58}).
The mixing-length is defined as $l\equiv \alpha H_{\rm p}$,
where $H_{\rm p}$ is the pressure scale height.
The convection zones are mixed via
a strong full mixing turbulent diffusion 
coefficient $d_{\rm fm}=\rm 10^{13}\,cm^2\,s^{-1}$ which produces a homogeneous
composition (Morel~\cite{m97}). 
As quoted in Sect.~\ref{sec:ipeg}, during the
pre-main sequence, we used a soft mixing of the convection zones. We found
empirically that the mixing generated by a turbulent diffusion coefficient
$d_{\rm mm}=\rm 10^6\,cm^2\,s^{-1}$
provides lithium depletions close to the observed values for both components.
We are aware that this procedure is not fully satisfactory: first, as already
quoted, we do
not know why and how the lithium destruction is annihilated in $\iota$\,Peg.\,B
during the pre-main sequence, second, we suppose {\it a priori} that the
observed depletion corresponds to the onset of the main sequence and, third,
we do not know how the unknown physical
process which acts on lithium performs on other chemicals. 

Following
the prescriptions of Schaller et al. (\cite{ssmm92})
the models take into account overshooting of mixed
convective cores over the distance $O_{\rm v}=0.2\min(H_{\rm p},R_{\rm co})$;
$R_{\rm co}$ is the radius of the convective core.

An atmosphere is restored using Hopf's $T(\tau)$ law (Mihalas \cite{m78}).
The connection with the envelope is made at the Rosseland optical
depth $\tau_{\rm b} = 10$, where the diffusion approximation for radiative
transfer becomes valid (Morel et al. \cite{m94}).
In the convective part of the atmosphere, a numerical trick
(Henyey et al. \cite{hvb65}) is employed in connection with the purely
radiative Hopf law in order to ensure the continuity of gradients
at the limit between the atmosphere and the envelope. 
At each time step, the radius $R_\star$ of the model
is taken at the optical depth $\tau_\star\simeq 0.6454$ where
$T(\tau_\star)=T_{\rm eff}$;
the mass of the star $M_\star$, is the mass inside the sphere of
radius $R_\star$.
The external boundary is located at the optical
depth $\tau_{\rm ext}=10^{-4}$, where the density is fixed to
$\rho(\tau_{\rm ext})=3.55\,10^{-9}$\,g\,cm$^{-3}$.

\subsection{Numerics.}
The models have been computed using the CESAM code (Morel \cite{m97}).
The numerical schemes are fully implicit and their accuracy
is first order for the time and third order for the space.
Typically, each model is described by about 600 mass shells,
it increases up to 2100 for the models used in seismological analysis. 

\begin{table}
\caption[]{
Characteristics of a calibration of the  $\iota$\,Peg. system.
$R_{\rm cz}$ and $R_{\rm co}$ are respectively the radius of the base of the
external convection zone and of convective core (including overshoot). At
the center,
$T_{\rm c}$, $\rho_{\rm c}$, $X_{\rm c}$, $Y_{\rm c}$ are respectively the
temperature (in M\,K), density (in g\,cm$^{-3}$),
hydrogen and helium mass fractions. 
}\label{tab:AB}
\begin{tabular}{llllllllll} \\  \hline \\
                     &$\iota$\,Peg.\,A&$\iota$\,Peg.\,B\\ \\ \hline \\  
$T_{\rm eff}$        &6642\,K        &4991\,K  \\
$L/L_\odot$          & 3.041         &0.3148 \\
$R/R_\odot$          & 1.320         &0.7499 \\
$R_{\rm cz}/R_\star$ &0.929          &0.696\\
$R_{\rm co}/R_\star$ &0.100          &0.166 \\
$T_{\rm c}$          & 17.21         &12.19\\
$\rho_{\rm c}$       & 86.83         &79.22\\
$X_{\rm c}$          &0.699          &0.682\\
$Y_{\rm c}$          &0.283          &0.297\\
$\rm [Li]$         &$+3.27$\,dex   &$+2.69$\,dex \\
\\ \hline \\
\end{tabular}
\end{table}

\begin{table}
\caption[]{
Calibration parameters of a $\iota$\,Peg. model lying within the uncertainty
boxes
and of a calibrated solar model using the same physics. The initial values are
labeled by ``$_0$''. The uncertainties are empirically estimated (see text).
}\label{tab:glob}
\begin{tabular}{llllllllll} \\  \hline \\
                &$\iota$\,Peg.     &Sun\\  \\ \hline \\
$\alpha_{\rm A}$&$1.46^{+0.20}_{-0.15}$   &1.76\\ \\
$\alpha_{\rm B}$&$1.36^{+0.30}_{-0.10}$\\ \\
$Y_0$           &$0.278^{+0.007}_{-0.007}$ &0.2722 \\ \\
age             & $0.00^{+500}_{-10}$\,My       &4.6\,Gy\\
 \\ \hline \\

$X_0$           &$0.705$ &0.7081  \\ 
$Z_0$           &$0.017$&0.0197\\ \\ \hline
\end{tabular}
\end{table}

\begin{figure*}
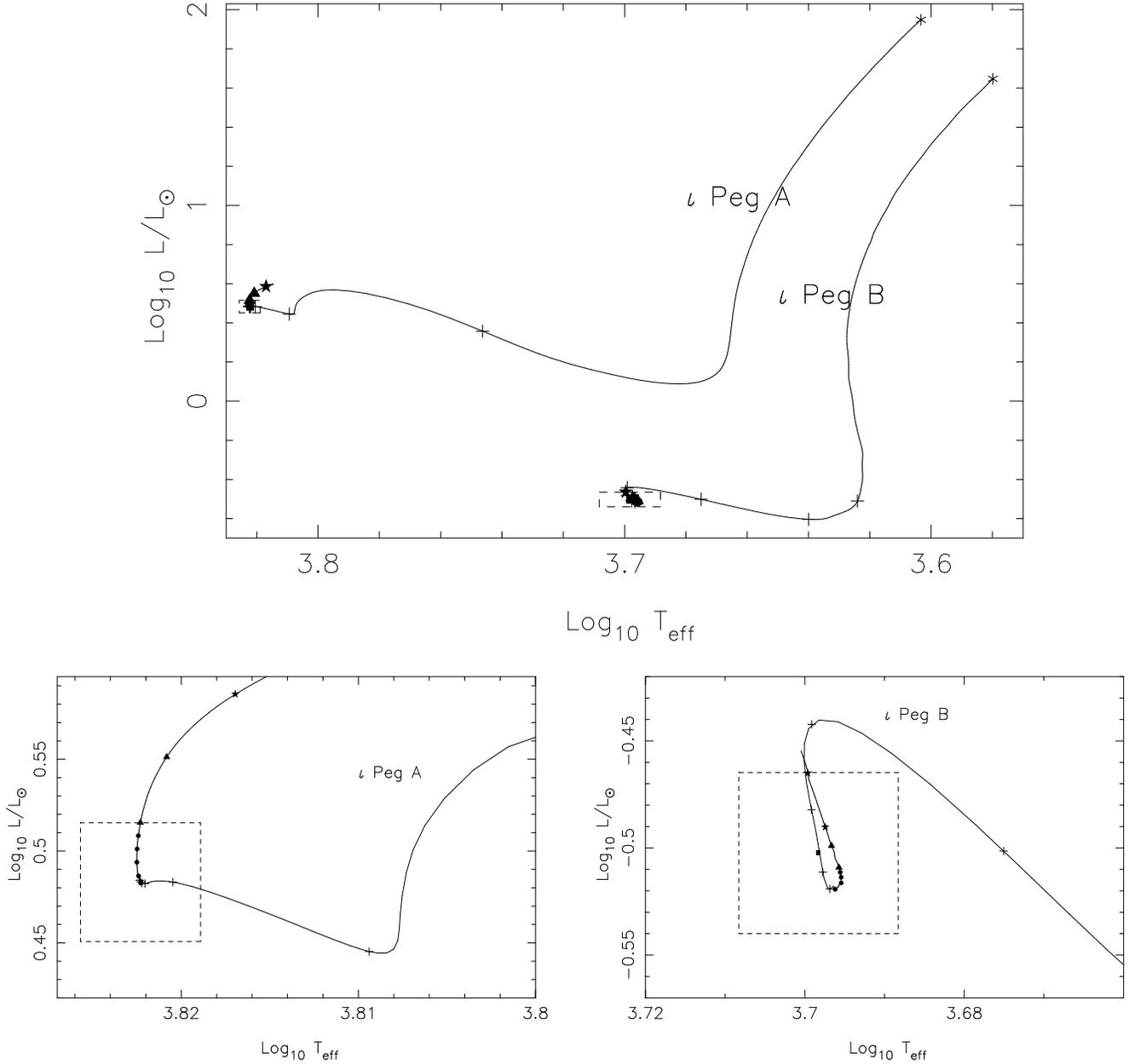

\vbox{
\centerline{
\psfig{figure=9046.f1.ps,height=10cm,angle=270}
}
\vspace{0.5truecm}
\hbox{\psfig{figure=9046.f2.ps,height=6cm,angle=270}
\hspace{0.5truecm}
\psfig{figure=9046.f3.ps,height=6cm,angle=270}}
}
\caption{
Top: Evolutionary tracks in the  HR diagram for $\iota$\,Peg.\,A and B.
Bottom: enlargements around the observational points (full square) with
the uncertainty domains. The stellar evolution sequences are
initialized (stick mark $\ast$) on the pre-main sequence
soon after the deuteron ignition.
Along the stellar tracks the time intervals between the marks
are respectively 10\,MY ($+$), 100\,MY ($\bullet$),
500\,MY (full triangle), 1500\,MY ($\star$). 
}\label{fig:AB}
\end{figure*}

\section{Results}\label{sec:res}
\subsection{Age and chemical composition}
In the HR diagram
Fig.~\ref{fig:AB} (bottom left) shows that the locus of $\iota$\,Peg.\,A
 belongs to its
uncertainty domain for evolutionary times in the interval
 $30\,{\rm My}\loa t_{\rm ev\,A}\loa 500\,{\rm
My}$. 
Effective temperature and luminosity are very close to their observed
values for $t_{\rm ev\,A}\simeq45$\,My.
The locus of $\iota$\,Peg.\,B (Fig.~\ref{fig:AB}, bottom right) belongs to its
uncertainty domain for times $45\,{\rm My}\loa t_{\rm ev\,B}\loa 3\,{\rm
Gy}$. The distances between the loci of theoretical effective
temperatures and luminosities and their observed values are smaller
for evolutionary times close to $t_{\rm ev\,B}\sim55$\,My
and $t_{\rm ev\,B}\sim1$\,Gy.

According to our definition in Sect.~\ref{sec:comp}, the ZAMS occurs
as soon as the nuclear dominates the gravitation as the primary source of
energy. That occurs at evolutionary
times $t_{\rm zams\,A}\simeq 36$\,My and $t_{\rm zams\,B}\simeq70$\,My for
$\iota$\,Peg.\,A and B, respectively. Therefore, the age
of the system $\iota$\,Peg.
is between $\sim$ ZAMS and $\sim0.5$\,Gy. For the calibrated models,
within our definition of the theoretical zero age main
sequence, the age of the binary system is about zero
i.e. $\iota$\,Peg.\,A and $\iota$\,Peg.\,B
are inhomogeneous ZAMS stars. The arguments in favor of this calibration 
are the following:
\begin{itemize}
\item the low amount of lithium depletion observed in the secondary is in
agreement with the observations,
\item in the HR diagram the loci of theoretical effective temperatures and
luminosities are close to the observations at
time $t_{\rm ev}\approx 56$\,My for
primary and secondary, while the differences are significantly
larger as soon as $\sim0.1\,{\rm Gy}\la t_{\rm ev}\la 0.5\,{\rm Gy}$,
\item the fact that the method of steepest descent converges
simultaneously for primary
and secondary models reveals a stable solution.
\end{itemize}

Table~\ref{tab:AB} shows the characteristics of
$\iota$\,Peg.\,A and B calibrated models.
The primary and the secondary have a convective core. The lithium depletion at
the surface of both components is close to the observed values.

Table~\ref{tab:glob} shows the calibration parameters obtained for
$\iota$\,Peg. and solar models computed with the same physics.
The uncertainties are derived empirically by
changing the calibration parameters around their values derived of the
calibrated model (the observed {\em solar} metallicity
of $\iota$\,Peg. is assumed to be error free).
Considering first, the non linearity of the solution
around ZAMS and second, the large
uncertainty domains, chiefly due to the inaccurate value of $\Delta m_{\rm
V}$, more elaborated estimations of calibration parameters uncertainties
via, e.g. Monte Carlo procedure, will need heavy computations, which are
not relevant of the scope of this paper.

\begin{table}
\caption[]{ 
 Theoretical global characteristics of the low degree p-mode spectrum 
of the star $\iota$\,Peg.\,A. All the quantities are given in $\mu$Hz (see 
text).
}\label{tab:seis}
\begin{tabular}{llllll} \\  \hline \\
   &age (My)&  $\nu_{0 0}$ &$\Delta\nu_0$ &  $\delta_{0,2}$  \\ 
   \\ \hline \\
A0 &50   &2234.91     &102.0          &  11.76            \\
A1 &150  &2222.57     &101.50         &  11.47            \\
A2 &500  &2106.61     &96.30          &  10.61            \\
\\ \hline \\
\end{tabular}
\end{table}
 
\subsection{Seismological analysis of $\iota$\,Peg.\,A}
The star $\iota$\,Peg.\,A belongs to the solar-like stars region of the HR 
diagram.
The oscillations of such stars may be stochastically excited by the convection, 
as it occurs in the Sun. The amplitudes of solar-like oscillations have been 
estimated by Houdek (\cite{h96}) for stars of different masses and ages. Here, 
we consider the oscillation spectrum in the frequency range where the 
predicted amplitude is maximum, i.e. between 1 and 3 mHz.

The oscillations of a non rotating star are characterized by two numbers:
the degree $\ell$ which is inversely proportional to the horizontal
 wavelength and the radial order $n$ (Christensen-Dalsgaard \& Berthomieu
 \cite{cb91}). Here the rotation is small, hence neglected. 
The frequencies which may be observed, according to Houdek, are on the 
low-degree ($\ell$=0 to 3) high-frequency p-modes (i.e. large radial order 
p-modes). The frequency spectrum is characterized by 
a few  global quantities, $\Delta\nu_0$ which describes the quasi-equidistance 
of the frequencies in the spectrum and $\delta_{\ell, \ell+2}$
the small frequency separations between frequencies of modes
with degrees of same parity and consecutive radial orders
$\nu_{n, \ell}$ and $\nu_{n-1, \ell+2}$.
They can be  derived  from   the following simplified polynomial 
expression (Berthomieu et al. \cite{b93}):
\[ \nu_{n, \ell} =  \nu_{0\ell} + \Delta\nu_0 (n+{\ell\over 2}-n_0).\] 
The quantity $\nu_{0\ell}$ varies little with the degree $\ell$.
We have taken $n_0$= 21, approximately in the middle of the range of the 
expected excited frequencies (i.e. $1 - 3$ mHz) according to Houdek
($loc.\,cit.$).
The large frequency separation $\Delta\nu_0$ is the distance between peaks of 
consecutive radial order for a given degree. It corresponds to the sound 
travel time across a stellar diameter and is mainly sensitive to the outer 
layers of the star.
The small frequency separations around $n=n_0$, defined by:
\[\delta_{\ell, \ell+2} = \nu_{n_0, \ell}-\nu_{n_0-1, \ell+2}
\sim \nu_{0\ell}-\nu_{0\ell+2},\]
are very sensitive to the core of the star.

Table~\ref{tab:seis} shows the quantities $\nu_{0,0}$, $\Delta\nu_0$ and 
$\delta_{0,2}\ (n_0=21)$ 
which  have been computed for three models  A0, A1, A2 of $\iota$\,Peg.\,A
in the observational uncertainty HR box, at different
evolutionary times 
$t_{\rm ev}=50,\,100,\,500$\,My.
These characteristic quantities 
 decrease with the age. $\nu_{00}$ and $\Delta\nu_0$
vary proportionally to $R^{-\frac32}$, 
as expected, since the structure of the three models are very similar.
The large separation $\Delta\nu_0$ which is the first quantity
to be derived by 
observations varies significantly with the age of the star.
The relative variation of $\delta_{0,2}$ is larger, due to the
differences of structure in the core, but more difficult to be observed.

\section{Discussion and conclusions}\label{sec:dis}
With the individual masses derived from spectroscopic
and visual orbits (Boden et al. \cite{bkb98}), we have attempted to calibrate the close binary system
$\iota$\,Peg. 
We have adopted effective temperatures and luminosities as derived from
photometric data with the K bolometric correction and the empirical scale of
temperature of Alonso et al. (\cite{aam95}, \cite{aam96}).
We have fixed the initial metallicity of primary and secondary to the
nowadays observed ({\em solar}) value. The evolutionary
sequences include the pre-main sequence. Microscopic diffusion and
gravitational settling of chemical species are allowed for.
To fit the observed lithium depletion in $\iota$\,Peg.\,B
we have used a mildly mixing of convection zones in the course of
the pre-main sequence. The convective cores are extended by $0.2\,H_{\rm p}$,
the standard amount of overshoot.

We have obtained adjustments within the uncertainty domains
for effective temperatures and
luminosities for both components with times of evolution within $\sim 40\,{\rm
My}\la t_{\rm ev}\la 0.5\,{\rm Gy}$. Owing to the lack of lithium depletion
observed in $\iota$\,Peg.\,B, a solution is obtained for the
evolutionary time
$t_{\rm \iota\,Peg.}\sim 56$\,My. The age of the system is then 
close to zero and we argue that the components of the binary
system $\iota$\,Peg. are non-homogeneous ZAMS stars.

Separate values are found for the mixing-length parameter of each component.
Even for $\iota$\,Peg.\,B, of mass smaller than solar,
Table~\ref{tab:AB} shows significant differences between the
convection parameter derived for each component and those
of the calibrated solar model
computed with the same physics.
This result appears in disagreement with the
universality of the solar convection parameter for
stellar masses between $0.6M_\odot\la M_\star\la 1.0M_\odot$
with metallicity close to solar (Fernandes et al. \cite{flbm98}).
Nevertheless, owing to the large observational
uncertainty domains, our result does not totally rule out the paradigm of the
universality of the solar convection parameter.

We find a larger mixing-length parameter for $\iota$\,Peg.\,A ($T_{\rm
eff}=6642$\,K) than for the cooler $\iota$\,Peg.\,B ($T_{\rm
eff}=4991$\,K). The calculations of
Ludwig et al. (\cite{lfs99}) seem to predict exactly the opposite.
We do not elucidate the reasons of this difference. We are aware
that our models are not strongly constrained by the observations. 
The largest uncertainty comes from the difficulty to estimate the magnitude
difference in the V band, a fundamental quantity for the determination of
luminosities and effective temperatures based on photometric data.
Improvements of the calibration of $\iota$\,Peg. will
necessitate a more accurate value for
the V magnitude difference.
Chemical composition determinations and estimates
of effective temperatures and gravities by spectroscopic 
detailed analysis of each component 
are perhaps realistic with modern technology (C. van\,'t
Veer, private communication) and are urgently needed.

We must also emphasize the fact that $\iota$\,Peg. is probably
tidally locked or, at least, recently unlocked 
and then it maybe a chromospherically active binary. Therefore to
model such a pair as two standard
isolated single stars is perhaps too simplistic.
In some years from now, one can expect that asteroseismology from space
e.g. COROT mission (Baglin \cite{bc98}) will provide more constraints on stellar
models as it is done nowadays with the Sun.
This preliminary work also shows the powerful interest of
interferometric and spectroscopic binaries such as $\iota$\,Peg. 
as tests of transport process theories in stellar modeling.

\begin{acknowledgements}
We would like to thank D. Bonneau, R. Cayrel, F. Th\'evenin,
C. and F. van\,'t Veer and J.P. Zahn for a number of helpful
discussions and comments, and
G. Cayrel de Strobel for valuable information on
metallicity. We are grateful to A. Baglin and Y. Lebreton for careful reading and
comments on the manuscript. Y. Lebreton
 kindly provides us the opacity package.
We thank the referee Dr. A.F. Boden for valuable and illuminating comments.
This research has made use of the Simbad data base, operated at
CDS, Strasbourg, France and of the General Catalogue of Photometric Data,
operated at the University of Lausanne, Switzerland.
This work has been performed using the computing facilities 
provided by the OCA program
``Simulations Interactives et Visualisation en Astronomie et M\'ecanique 
(SIVAM)''.
\end{acknowledgements}

\end{document}